\documentclass[%
 reprint, 
 superscriptaddress,
 amsmath,amssymb,
 aps,
floatfix,
longbibliography
]{revtex4-1}

\usepackage{graphicx}
\usepackage{bm}

\usepackage{xcolor}
\usepackage[colorlinks=true,citecolor=blue,linkcolor=magenta]{hyperref}

\newcommand{\ben}{\begin{eqnarray}}
\newcommand{\een}{\end{eqnarray}}

\newcommand{\be}{\begin{equation}}
\newcommand{\ee}{\end{equation}}

\newcommand{\s}{\scriptscriptstyle}

\begin{document}

\title{Effect of decay of the final states on  the probabilities of the Landau-Zener transitions in multistate non-integrable models}

\author{Rajesh K. Malla}
\affiliation{Center for Nonlinear Studies and Theoretical Division, Los Alamos National Laboratory, Los Alamos, New Mexico 87545, USA}
\author{M. E. Raikh}
\email{raikh@physics.utah.edu}
\affiliation{Department of Physics and
Astronomy, University of Utah, Salt Lake City, UT 84112}

\begin{abstract}
For a Landau-Zener transition in a two-level system, the probability for a particle, initially in the first level, to survive the transition 
and to remain in the first level, does not depend on whether or not the second level is broadened  [V. M. Akulin and W. P. Schleicht,  Phys. Rev. A {\bf 46}, 4110 (1992)]. In other words, the seminal Landau-Zener result applies regardless of the broadening of the second level. The same question for the multistate Landau-Zener transition is addressed in the present paper. While for integrable multistate models, where the transition does not involve interference of the virtual paths, it can be argued that the independence of the
broadening persists, we focus on non-integrable models involving interference. For a simple four-state model, which allows an analytical treatment, we demonstrate that the decay of the excited states affects the survival probability provided that {\em the widths of the final states are different}.
\end{abstract}

\maketitle

\section{Introduction}

In Ref. \onlinecite{Akulin1992}
(see also Ref. \onlinecite{Vitanov1997}) a problem of time-dependent crossing between a sharp level and a broadened level was considered.  It was assumed that, initially, at times $t\rightarrow -\infty$, the sharp level is the ground state and is occupied. The authors 
demonstrated that the probability for a sharp level to stay occupied after the transition, at $t\rightarrow \infty$, does not depend on the width of the broadened level. The authors of Ref. \onlinecite{Akulin1992} had arrived at their conclusion by a direct calculation although it could be argued on general grounds (see e.g. Ref. \onlinecite{Shytov}) that the level width drops out from the  $t\rightarrow \infty$ limit of the survival probability since this probability emerges as a result of integration along the contour of a large radius in the complex time plane. 

A natural question to ask is whether the decay of the excited states affects the survival probability in the case of multilevel transitions. These transitions are characterized by $N\times N$ scattering matrix with $N\geq 3$. Finding the elements of this matrix analytically is possible only for certain mutual arrangements of the levels, \cite{Demkov1968,Brunhobler,Ostrovsky1997,Ostrovsky2000,Ostrovsky2001,Sinitsyn2004,Sinitsyn1,Volkov}.  
Recently\cite{Sinitsyn2015,Sinitsyn2017,Yuzbashyan2017,Sinitsyn2018,Sinitsyn2020} a new class of exactly solvable multistate models was uncovered and the conditions under which the problem is solvable were formulated. A common feature of the solvable models is that the elements of the scattering matrix are expressed via partial Landau-Zener probabilities and do not depend on the time intervals between the subsequent level crossings. This fact is crucial for the effect of the level decay on the survival probability. For exactly solvable (integrable) models the levels' decay should drop out from the survival probabilities. On the contrary, for non-integrable multistate models, it might be expected that the probabilities should depend on the ratio of the time intervals between the level crossing and the lifetimes of the crossing levels. In the present paper, we consider a non-integrable four-state model for which the consideration can be carried out analytically. By including the broadening of the two possible final states of the transition, we trace how this broadening modifies the survival probability.
\vspace{4mm}

\section{Model}

For illustration purposes, we consider two tunnel-coupled quantum dots in which the levels are swept past each other by the gate voltage, see Fig. 1.  Denote with $v$ the relative velocity of the levels and with $J$ the tunnel
splitting of the levels at the crossing point. In order to mimic the decay, we assume that the electron can tunnel out of the dot.  This tunneling causes an imaginary correction,
$\frac{i}{\tau}$, to the energy level position, $\frac{vt}{2}$, in the left dot, where $\tau$ is the tunnel  lifetime. Then the amplitudes $a$ and $b$ to find a particle in the left and right dot, respectively, are related via the Schr{\"o}dinger equation
\begin{align}
	\label{a}	
	&i\left({\dot a}+\frac{a}{\tau}\right)=-\frac{vt}{2}a+Jb,\\
	&\label{b}
	i{\dot b}=\frac{vt}{2}b+Ja.
\end{align}
The solutions of the system Eqs. (\ref{a}), (\ref{b}) is governed by two dimensionless parameters: $\frac{J^2}{v}$ and $\frac{J}{v\tau}$.
Then the result of Refs. \onlinecite{Akulin1992}, \onlinecite{Vitanov1997} reduces to
the fact that the celebrated Landau-Zener formula 
\begin{equation}
	\label{celebrated}	
	\frac{|b({\infty})|^2}{|b(-\infty)|^2}=\exp\left( 	 -\frac{2\pi J^2}{v} \right).         
\end{equation} 
applies independently of the parameter $\frac{J}{v\tau}$.

A hint on how the decay in the left dot affects  the survival probability can be inferred from the semiclassical solution
of the system Eqs. \ref{a} and \ref{b}. Upon substituting
$a(t)=a_0e^{-i\lambda t},~~  b(t)=b_0e^{-i\lambda t}$,
we arrive at the system of algebraic equations
\begin{align}
\label{system}
&\left(\lambda +\frac{i}{\tau}   +\frac{vt}{2} \right)a_0=Jb_0\nonumber\\
&\left(\lambda -\frac{vt}{2}    \right)b_0=Ja_0.	
\end{align}
This system yields the following equation for $\lambda$
\begin{equation}
\label{lambda}
\left(\lambda +\frac{i}{2\tau} \right)^2=
\left( \frac{vt}{2}  +  \frac{i}{2\tau} \right)^2 +J^2.	
\end{equation}
At large times the term $\frac{i}{2\tau}$ is negligible compared to $\frac{vt}{2}$. Since, according to the contour integral approach 
\cite{Demkov1968,Shytov,Sinitsyn2004}, the survival probability is fully determined by the behavior of $a(t)$ and $b(t)$  at large times, it can be expected that the decay does not affect this probability. In the absence of decay, the quantitative criterion of large times is $t\gg \frac{J}{v} $. In the presence of the decay, there is an additional criterion  of large times is 
$t\gg \frac{1}{v\tau} $.  The physical meaning of the latter criterion is that the change,  $vt$, of the level positions during the time, $t$, must exceed the level width. The above consideration, however,  does not capture the behavior of the population of a narrow level,  $|b(t)|^2$, at finite times. We address this question in the next section.

\begin{figure}
\label{F1}
\includegraphics[scale=0.35]{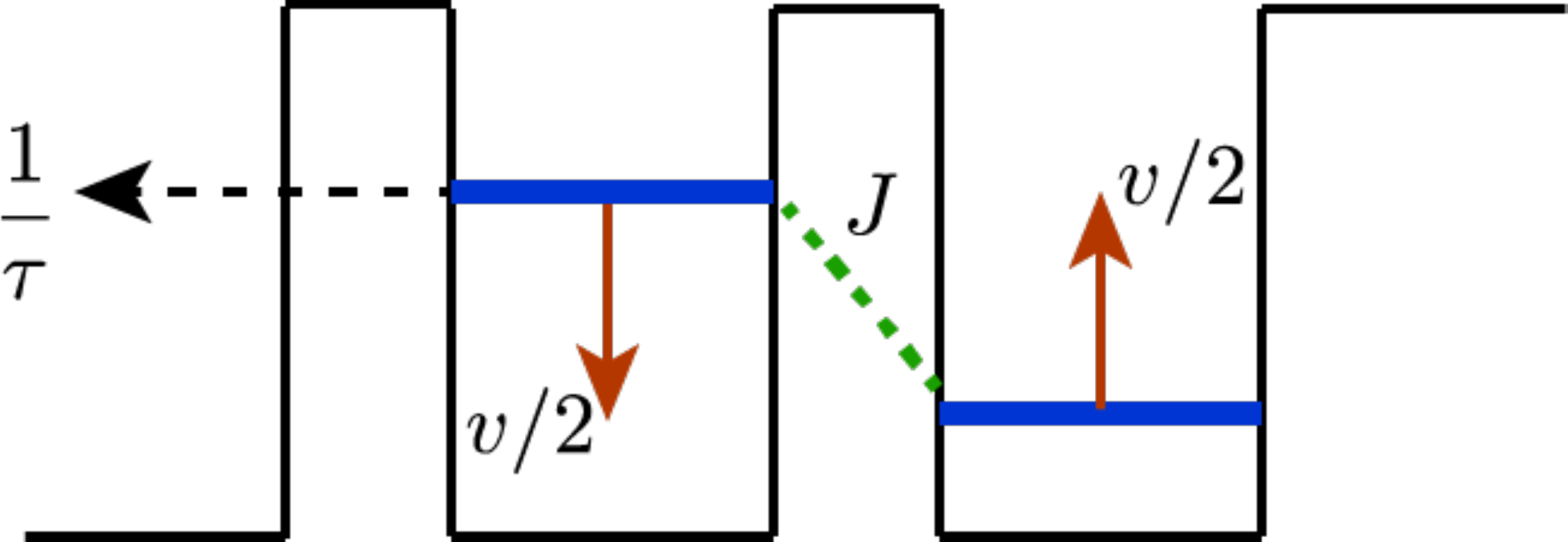}
\caption{(Color online) Microscopic illustration of the Landau-Zener transition with decay.	Shown  are two quantum 
dots coupled by tunneling with a matrix element,
$J$. Due to the temporal change of the gate voltage,
the levels in the dots are swept past each 
other with relative velocity, $v$. The left dot is
coupled to the continuum; the possibility of tunneling out
leads to the broadening,  $\frac{1}{\tau}$,    of the level in the left dot. At time $t\rightarrow -\infty$ an electron resides in the right dot.
Survival probability is the probability for the
electron to stay in the right dot at  $t\rightarrow \infty$.}
\end{figure}

\section{Effect of decay at finite times}

We start by reproducing more rigorously the $t\rightarrow \infty$ result of Refs. \onlinecite{Akulin1992}, \onlinecite{Vitanov1997} and then turn to finite $t$. 
To this end,  we express $b$ from  Eq.~\ref{a} and substitute it into Eq.~\ref{b}. Searching for $b(t)$ in the form $b_0(t)\exp(-\frac{t}{2\tau})$, we find that $b_0(t)$ satisfies the second-order differential equation
\begin{equation}
	\label{second-order}
	{\ddot b_0}	+\Biggl[\frac{1}{4}\left(vt+\frac{i}{\tau}    \right)^2+\frac{iv}{2}+J^2 \Biggr]b_0=0,
\end{equation}	
which is the equation for the parabolic cylinder functions.
Disappearance of $\tau$ from the probability, $\vert b(t)\vert^2$, in the long-time limit follows from the fact that in this limit the semiclassical
approach applies. The semiclassical phase is given by
$\Phi(t)=\pm i\int\limits_0^t dt
\biggl\{\frac{1}{4}\left(vt+\frac{i}{\tau}    \right)^2+\frac{iv}{2}+J^2 \biggr\}^{1/2}$. Keeping only two leading terms, in the long-time limit, we have 
$\Phi(t)\approx \pm i\left(\frac{vt^2}{4} +
\frac{it}{2\tau}  \right)$. 
The sign "-" leads to a growing exponent
$\propto \exp\left(\frac{t}{2\tau}  \right)$ in $b_0(t)$. This term is canceled by the prefactor $\exp\left(-\frac{t}{2\tau}\right)$ in $b(t)$. This cancellation  illustrates the message of Refs. \onlinecite{Akulin1992}, \onlinecite{Vitanov1997} that the broadening drops out from the survival probability at $t\rightarrow \infty$.{\tiny }
However, as the numerics in Ref. \onlinecite{Vitanov1997} shows, at {\em finite times}, the oscillations in $|b(t)|^2$ are suppressed  due to the leakage from the left dot into continuum. To trace this suppression analytically, 
we examine the solution of Eq.~ (\ref{second-order}) at complex argument 
$t+\frac{i}{v\tau}$.

At $\tau \rightarrow \infty$, i.e. without the decay, the solution of Eq.~ (\ref{second-order}), which satisfies the right ``boundary" condition $\bigl[
\exp(\frac{ivt^2}{4})$   behavior at $t\rightarrow \infty\bigr]$, is given by the function $D_{\nu}(z)$ with integral representation
\begin{equation}
	\label{representation}	
	D_{\nu}(z)=\Bigl(\frac{2}{\pi} \Bigr)^{1/2}e^{\frac{z^2}{4}}\int\limits_0^{\infty}	du~
	e^{-\frac{u^2}{2}}u^{\nu}\cos\left(zu-
	\frac{\pi\nu}{2}  \right).
\end{equation}

Argument $z$ and parameter $\nu$
in Eq. \ref{representation}  are defined as
\begin{equation}
	\label{defined}	
	z=v^{1/2}e^{\frac{\pi i}{4}}t,~~~ \nu=-\frac{iJ^2}{v}.
\end{equation}
Following Ref. \onlinecite{Zhuxi}, it is convenient to divide
the integral Eq. (\ref {representation}) into two contributions
\begin{equation}
	\label{two}
	D_{\nu}(z)=	I_{+}(z)e^{\frac{i\pi\nu}{2}} +
	I_{-}(z)e^{-\frac{i\pi\nu}{2}},
\end{equation}
where the functions $I_+(z)$ and $I_-(z)$ are defined as
\begin{equation}
\label{definition}
I_\pm(z)=	D_{\nu}(z)=\Bigl(\frac{1}{2\pi} \Bigr)^{1/2}e^{\frac{z^2}{4}}\int\limits_0^{\infty}	du~
u^{\nu}e^{-\frac{u^2}{2}\pm iuz}.
\end{equation} 
Both integrals are evaluated with the help
of steepest descent approach. The corresponding saddle points are given by
\begin{equation}	
	\label{saddle}
	u_{\pm}=\frac{iz}{2} \pm \frac{(4\nu-z^2)^{1/2}}{2}.
\end{equation}
Performing the Gaussian integration around the saddle points, we arrive to the following asymptotic expressions for $I_+$ and $I_-$
\begin{align}
\label{I+}
&I_+(z)\approx (4\nu-z^2)^{-1/4} \exp\Biggl[\frac{1}{4}iz(4\nu-z^2)^{1/2}  -\frac{1}{2}\nu  \Biggr]\nonumber\\
&\times	\Biggl[\frac{1}{2}iz+\frac{1}{2}(4\nu-z^2)^{1/2} \Biggr]^{\nu+\frac{1}{2}},
\end{align}
	
\begin{align}
	\label{I-}
	&I_-(z)\approx (4\nu-z^2)^{-1/4} \exp\Biggl[-\frac{1}{4}iz(4\nu-z^2)^{1/2}  -\frac{1}{2}\nu  \Biggr]\nonumber\\
	&\times	\Biggl[-\frac{1}{2}iz+\frac{1}{2}(4\nu-z^2)^{1/2} \Biggr]^{\nu+\frac{1}{2}}.
\end{align}	
The condition of applicability of the steepest descent method is that the typical value of   $\left(u-u_+\right)$
contributing to the integral Eq. \ref{definition}
is much smaller than $u_+$. 
With characteristic time of the transition being
$\frac{J}{v}$, we conclude that characteristic $z$ is $\sim \frac{J}{v^{1/2}}\sim |\nu|^{1/2}$.  Thus, for $|\nu|\gg 1$, when the position of the saddle point is $u_+ \sim |\nu|^{1/2} \gg 1$ the saddle-point result applies not only for large, but
for arbitrary $z$.\cite{withEugene}

 It  is instructive to rewrite $D_{\nu}(z)$ in terms of time

\begin{align}
\label{INTERMS}	
&D_{\nu}(z)=e^{\frac{\pi i}{8}+\frac{iJ^2}{2v}}\biggl(\frac{v}
{W^2(t)}   \biggr)^{1/4}
\Biggl(\frac{e^{-\frac{\pi i}{4}}}{v^{1/2}}\Biggr)^{\frac{1}{2}-
	\frac{iJ^2}{v}}
\nonumber\\
&\times \Bigg\{e^{{\frac{it}{4}}W(t)}\Bigg(\frac{-vt+
	W(t)}{2}
\Bigg)^{\frac{1}{2}-\frac{iJ^2}{v}}
e^{-\frac{\pi J^2}{2v}}
\nonumber\\
&+e^{{-\frac{it}{4}}W(t)}\Bigg(\frac{vt+
	W(t)}{2}
\Bigg)^{\frac{1}{2}-\frac{iJ^2}{v}}
e^{\frac{\pi J^2}{2v}}    \Bigg\},	
\end{align}
where $W(t)=\left(4J^2+v^2t^2\right)^{1/2}$.
Interference of two oscillating exponents in
the curly brackets
is responsible for the beatings in $\vert b_0(t)\vert^2$. The magnitudes of these terms are related as $\exp(-\frac{\pi J^2}{v})$, 
in agreement with classical result Eq. \ref{celebrated}.	
In the limit $t\rightarrow 
\pm \infty$ we have $W(t)\rightarrow \pm vt$, so either $\bigl(W(t)-vt\bigr)$ or $\bigl(W(t)+vt\bigr)$ approaches zero. In other words, only one of the two terms in the curly brackets survives in both limits. Moreover, the product of the two terms in the curly brackets does not depend on time.

Upon substituting $t+\frac{i}{v\tau}$ into
$\exp\Big(\frac{itW(t)}{4}\Big)$ and~~~~$\exp\Big(-\frac{itW(t)}{4}\Big)$, we realize that the $\tau$-dependence is canceled by the  prefactor $\exp\left(-\frac{t}{2\tau}\right)$ both at large positive and large negative $t$. Still, at finite $t$, the occupation of the right dot, in which the particle initially (at $t\rightarrow -\infty$) resides, depends on $\tau$. Namely, it is multiplied
by $\exp\Big(-2{\cal F}(t)  \Big)$, where the function ${\cal F}(t)$ is defined as

\begin{equation}
\label{calF}
{\cal F}(t)=\frac{\tau_{\s LZ}}{\tau}
\Biggl\{\frac{8}{(4+{\tilde t}^2)^{1/2}\left[\vert{\tilde t}\vert+(4+{\tilde t}^2)^{1/2}\right]^2}   \Biggr\},
\end{equation}
where ${\tilde t}=\frac{t}{\tau_{\s LZ}}$ is the 
dimensionless time measured in the units of 
$\tau_{\s LZ}=\frac{J}{v}$, which is the
characteristic time of the Landau-Zener transition.

We now turn to the interference effects in the 
level occupation. This occupation is proportional to
$\exp\left(-\frac{t}{\tau}  \right)|b_0(t)|^2$.
The essential time dependence of $b_0$ is 
contained in the curly brackets in Eq. \ref{INTERMS}. Upon taking the absolute value square of this curly brackets, we obtain	
\begin{align}
\label{interference}
&\Big\vert\Big\{ 	...  \Big\}   \Big\vert^2
=\exp\Big(-\frac{\pi J^2}{v}\Big)F+
\exp\Big(\frac{\pi J^2}{v}\Big)G\nonumber\\
&+2\left(FG\right)^{1/2}
\cos\Big(\frac{t}{2}W(t)-\frac{J^2}{v}\ln\frac{F}{G}\Big),
\end{align}
where the functions $F(t)$ and $G(t)$ are defined as
\begin{equation}
	\label{CandD}
	F(t)=\frac{W(t)-vt}{2},~~~G(t)=\frac{W(t)+vt}{2},
\end{equation}
so that the product $F(t)G(t)$ is equal to $J^2$, i.e. it
is time-independent. We then conclude that the time decay of the interference term is 
$\exp\big(-\frac{t}{\tau}\big)$. This dependence comes from the 
prefactor.

\vspace{4mm}

\vspace{4mm}

\begin{figure}
	\label{F2}
	\includegraphics[scale=0.3]{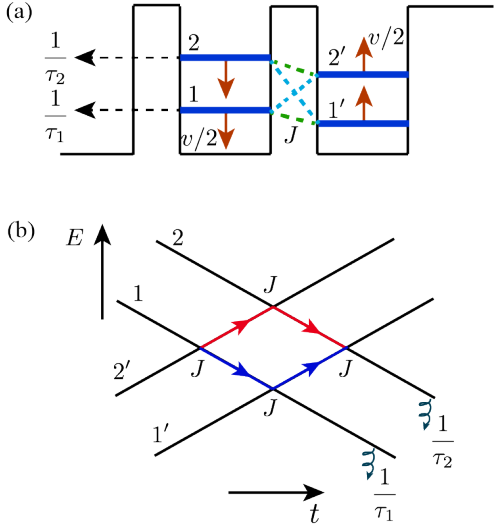}
	\caption{(Color online)
(a) Illustration of $2\times 2$ model with decay. Both left and right dots contain two levels. We assume that the level separation in both dots is
the same. It is also assumed that the coupling constants between the levels $1$ and $2$ in the left dot and the levels $1'$ and $2'$ in the right dot are the same and are equal to $J$. Similar to Fig. 1, the levels $1', 2'$ are swept by the levels $1, 2$ with relative velocity, $v$. Levels in the left dot are coupled to the continuum. Tunnel lifetimes into continuum, $\tau_1$ and $\tau_2$, of the levels $1$ and $2$ are {\em different.} As a result of this difference, the survival probability in the right dot depends on the decay from the right dot. (b) Time evolution of the levels
in the left and right dot. Unlike the
conventional Landau-Zener transition,
illustrated in Fig. 1, there are {\em two} pathways (red and blue) from the states in the right dot into the states in the left dot, which interfere with each other.}
\end{figure}

\section{$2\times 2$ non-integrable model with decay}

We now turn to the simplest non-integrable model illustrated in Fig. 2.
%{\bf FIG}.
In this model, there are two sharp levels in the right dot, which are spaced in energy by $2\Delta$, and two broadened levels in the left dot, which are coupled to the continuum. For simplicity, we assume that the level separation in the left dot is also  $2\Delta$. Also,
to keep the analysis as simple as possible, we assume that the tunnel couplings between all the levels are the
same and are equal to $J$.  Similar to the situation 
considered in the previous sections, we assume that the
levels in the left and right dots are swept past each other with velocity $v$. We denote the amplitudes to 
find the particle in the level $1$ and $2$ of the left
dot with $a_1$ and $a_2$, respectively. Correspondingly, $b_1$ and $b_2$ are the amplitudes to find the particle in the first level 
(level 1' in Fig. 2 and the second level (level 2' in the right dot. 
In the semiclassical limit, four amplitudes are related by the 
following system of  equations similar to
the system Eq. \ref{system}
\begin{align}
	\label{1}	
	&\left(\lambda+\frac{i}{\tau_{\s 1}}\right)a_1=-\left(\Delta+
	\frac{vt}{2}\right)a_1+J\left(b_1+b_2 \right),\\
	\label{2} 	
	& \left(\lambda+\frac{i}{\tau_{\s 2}}\right)a_2=\left(\Delta-
	\frac{vt}{2}\right)a_2+J\left(b_1+b_2  \right),\\
	\label{3}
	&\lambda b_1=\left(-\Delta+
	\frac{vt}{2}\right)b_1+J\left(a_1+a_2  \right),\\
	\label{4}
	&\lambda b_2=\left(\Delta+
	\frac{vt}{2}\right)b_2+J\left(a_1+a_2  \right).
\end{align}
Here $\tau_1$ and $\tau_2$ are the tunnel lifetimes of the levels in the left dot, and $\lambda$ is the time-dependent eigenvalue.
Introducing the sums and differences
\begin{align}
	\label{new1}	
	&A_1=a_1+a_2,~~~~~~~B_1=b_1+b_2,\\
	\label{new2}
	&A_2=a_1-a_2,~~~~~~~B_2=b_1-b_2,	
\end{align}
The system of four equations for the new variables is obtained 
by adding Eqs. (\ref {1}) and  (\ref {2})  and also by adding 
Eqs. (\ref {3}) and  (\ref {4}). This yields

\begin{align}
	\label{+}
	&\left(\lambda+\frac{vt}{2}+\frac{i}{2\tau_{\s 1}}+\frac{i}{2\tau_{\s 2}} \right)A_1-2JB_1\nonumber\\
	&=\left(-\Delta -\frac{i}{2\tau_{\s 1}}+\frac{i}{2\tau_{\s 2}}   \right)A_2,\\
	\label{++}
	&\left(\lambda-\frac{vt}{2} \right)B_1-2JA_1=-\Delta B_2.
\end{align}
The other two equations emerge upon subtracting  Eq. (\ref {2}) from Eq. (\ref {1}) and Eq. (\ref {4}) from Eq. (\ref {3})
\begin{align}
	\label{-}
	&\left(\lambda+\frac{vt}{2}+\frac{i}{2\tau_{\s 1}}+\frac{i}{2\tau_{\s 2}} \right)A_2=-\left(\Delta +\frac{i}{2\tau_{\s 1}}-\frac{i}{2\tau_{\s 2}}   \right)A_1,\\
	\label{--}
	&\left(\lambda-\frac{vt}{2} \right)B_2=-\Delta B_1.
\end{align}
In the limit of large $J$ we can neglect the first terms
in Eqs. (\ref {+}) and (\ref {++}) and express $B_1$ via $A_2$
and $A_1$ via $B_2$. This yields
\begin{align}
	\label{shortened}
	&B_1=\frac{\left(\Delta+\frac{i}{2\tau_{\s 1}}-\frac{i}{2\tau_{\s 2}}  \right)  }{2J}A_2,\nonumber\\
	&A_1=\frac{\Delta}{2J}B_2.
\end{align} 
As a last step, we substitute $B_1$ and $A_1$ into the right-hand sides of Eqs. (\ref{-}) and (\ref{--}), respectively, and arrive to
the final system
\begin{align}
	\label{final1}
	&\left(\lambda-\frac{vt}{2}\right)B_2=-\frac{\Delta\left(\Delta+     \frac{i}{2\tau_{\s 1}}-\frac{i}{2\tau_{\s 2}} \right)}
	{2J}A_2,\\
	\label{final2}
	&\left(\lambda+\frac{vt}{2}+\frac{i}{2\tau_{\s 1}}+\frac{i}{2\tau_{\s 2}} \right)A_2=-\frac{\Delta\left(\Delta+     \frac{i}{2\tau_{\s 1}}-\frac{i}{2\tau_{\s 2}} \right)}
	{2J}B_2.
\end{align}
Multiplying Eqs. (\ref{final1}), (\ref{final2}), we arrive to the closed
equation for $\lambda(t)$
\begin{equation}
	\label{closed}
	\left(\lambda(t)+\frac{i}{4\tau_{\s 0}}\right)^2=
	\left(\frac{vt}{2}+  \frac{i}{4\tau_{\s 0}}\right)^2 +\frac{\Delta^2\left(  \Delta+\frac{i}{2\tau_{\s c}}\right)^2}{4J^2},	
\end{equation}	
where $\tau_0$ and $\tau_c$ are defined as
\begin{equation}
	\label{Defined}
	\frac{1}{\tau_{\s 0}}=
	\frac{1}{\tau_{\s 1}}+\frac{1}{\tau_{\s 2}},~~~~~~\frac{1}{\tau_{\s c}}=
	\frac{1}{\tau_{\s 1}}-\frac{1}{\tau_{\s 2}}.	
\end{equation}
Comparing Eq. \ref{lambda} to  Eq. \ref{closed}, we see
that the semiclassical description of the Landau-Zener transition in $2\times 2$ model reduces to the conventional 
Landau-Zener theory in which the tunnel splitting $J$ is
replaced by
\begin{equation}
\label{replaced}
{\tilde J}=\frac{\Delta\left(\Delta+\frac{i}{2\tau_c}\right)}{2J}={\tilde J}_1+i{\tilde J}_2.
\end{equation}   
Naturally, in the absence of decay $\tau_c\rightarrow
\infty$, Eq.~\ref{replaced} reproduces the earlier
result obtained in Ref. \onlinecite{Rajesh}
together with a  nontrivial  message that 
the survival probability {\em increases}
upon the increasing of the tunnel matrix element, $J$.
Later this message was confirmed by a semiclassical
calculation in Ref. \onlinecite{Rajesh1}.

Qualitative  novelty of the system Eqs. \ref{final1},
\ref{final2} is that the four-level $2\times 2$ model is reduced to the conventional Landau-Zener
system Eq.~\ref{system} with an effective tunnel matrix element,
${\tilde J}$, being {\em complex}. This complexity 
reflects the effect of decay  in generic non-integrable models.

A straightforward consequence of the effective tunneling matrix element being complex is that 
the dimensionless parameter defined in Eq. \ref{defined} becomes  {\em complex}

\begin{equation}
\label{complex}
\nu =\nu_1+i\nu_2=-\frac{i{\tilde J}^2}{v} =-\frac{\Delta^3}{2J^2\tau_cv}-
\frac{i\Delta^2\left(\Delta^2-\frac{1}{4\tau_c^2}\right)   }{4J^2v}.
	\end{equation}
In the absence of decay, the survival probability emerges 
as an absolute value square of the ratio of the long-time asymptotes of the two terms in Eq. \ref{two}. With parameter $\nu$ being complex,
upon taking the absolute value square, the real
part of $\nu$ drops out. This leads us to the
main result: 
%Our main result is that 
the survival
probability in a non-integrable $2\times 2$ model illustrated in Fig. 2 is given by
\begin{equation}
\label{main}
\exp{\Big(-2\pi|\nu_2|  \Big)}=
\exp{\Bigg[-\frac{2\pi\Delta^2
\Big{\vert}\Delta^2-\frac{1}{4\tau_c^2}\Big{\vert}}{J^2v}    \Bigg]}.
\end{equation}	
At first glance, this result seems counterintuitive.  Indeed, Eq. \ref{main} suggests 
a nontrivial dependence of the survival probability on the decay, which is peaked at  
$\frac{1}{|\tau_c|}=2\Delta$.
On the other
hand, the decay enters in combination with a half of inter-level energy separation,  which is intuitively correct. 
As explained in Ref. \onlinecite{Rajesh}, the inverse dependence of the survival probability on the tunnel splitting is the result of the interference of the multiple tunnel paths and is the consequence of the $2\times 2$ model being non-integrable. 

\vspace{5mm}

\section{Relation to Ref. [20]}

In general, the transition probabilities in multi-state 
Landau-Zener models cannot be evaluated analytically.  
This is due to the virtual paths interference that emerges  when multiple paths lead up to a particular transition.
This interference is illustrated in Fig. 2b.  The systems of general type with finite path interference are not fully solvable. In Ref. \onlinecite{Rajesh} such not-fully solvable models were referred to as non-integrable. 
However, 
%according to  
in Ref.~\onlinecite{Rajesh1}
%Ref. \onlinecite{commute} 
%Ref. \onlinecite{Rajesh1}
integrability was defined in a different sense.
Namely, the integrability condition for a time-dependent Hamiltonian $H(t,t')$ was defined as a possibility to find analytical form of a nontrivial operator $H'(t,t')$ such that \cite{commute}
\begin{eqnarray}
	\label{cond1}
	\frac{\partial H}{\partial t'} - \frac{\partial H'}{\partial t} &=&0,\\
	\label{cond2}
	[H,H']&=&0.
\end{eqnarray} 
In particular, in Ref. \onlinecite{Rajesh1} it was demonstrated that the Landu-Zener grids, with two sets of parallel levels, similar to Fig. 1a, are formally integrable.   These grids have non-trivial commuting partners having the form of time-quadratic-polynomial operators. While these models are still not fully solvable,
integrability in the sense of Ref. \onlinecite{Rajesh1}  allows
to establish exact non-trivial relations between the partial 
transition probabilities. It also allows finding asymptotically exact expressions for the leading exponents describing the transition 
probabilities in the nearly adiabatic limit.

In Ref. \onlinecite{Rajesh1} general theoretical predictions  were put
 to numerical test  for a linearly driven double quantum dot system, %where the system is 
 described by a 
 Landau-Zener grid with two sets of parallel levels crossing each other, see Fig. 2a.

% The above hypothesis about integrability was put to the test for a %linearly driven double quantum dot system, where the system is %described by a 
%Landau-Zener grid with two sets of the parallel levels crossing each %other. 

Within the approach of Ref. \onlinecite{Rajesh1}, the time-dependent Hamiltonian is described by a $4\times4$ matrix. Integrability reduces the system to an effective $2\times 2$ matrix Hamiltonian, and the transition probabilities are obtained by a modified Dykhne formula. Most importantly, the strong coupling limit of Ref. \onlinecite{Rajesh1}, similar to the limit $\Delta \ll J$ in 
Section IV, the numerical test reproduced the result of Ref. 
\onlinecite{Rajesh}.

%There were four parameters including,  level spacings between two %levels in quantum dot 1, $e_1$, and quantum dot 2, $e_2$, coupling %between diabatic levels, $g$, and the difference between the slopes of %two quantum dots $b$. The semiclassical solution is applicable in a %broad range of parameter regimes and shows two topologically distinct %phases for $e_1e_2/g^2\ll 1$ and $e_1e_2/g^2>1$.  We found that in the %limit  $e_1e_2/g^2\ll 1$ and $g^2/b\gg1$ this result agrees with the %result previously obtained in Ref. \onlinecite{Rajesh},

In the presence of decay, the Hamiltonian representing the system of equations (\ref{1}), (\ref{2}), (\ref{3}), (\ref{4}) is non-Hermitian. Nevertheless, this Hamiltonian satisfies the integrability conditions (\ref{cond1}) and (\ref{cond2}), where the commuting partner operator is also non-Hermitian. The integrability allows for rescaling of the parameters by changing $t'$ as outlined in Ref. \onlinecite{Rajesh1} and could lead to a semiclassical solution even in the presence of decaying terms. This is however beyond the scope of the present paper.

\section{Concluding remarks}  
In the present paper, we demonstrated for the first time that, unlike the case of integrable multistate models, where the decay into the continuum does not affect the survival probability, in non-integrable models, this decay enters explicitly into the survival probability, see Eq. \ref{main}. The model considered in the
present paper is the simplest example of the
Landau-Zener grid.\cite{Rajesh1,Suzuki2022}
Note that, for decay to enter the result of the transition, the tunnel widths of the two final states {\em should be different}, as it follows from Eq. \ref{Defined}.   	
The underlying reason is that the imbalance of the
tunnel widths results in the mixing of the symmetric and antisymmetric modes in the two dots. A crucial step that allowed us to solve the 
system \ref{1} is the large-$J$ limit. Quantitatively, this  large-$J$ limit corresponds to the condition ${\tilde J}\ll J$, where ${\tilde J}$
is defined by Eq. \ref{replaced}. As follows from Eq. \ref{replaced},
this condition amounts to $|\Delta|\ll J$ and 
$\frac{|\Delta|}{\tau_c}\ll J^2$. It also follows from 
Eq.~\ref{shortened} that the smallness of the ratio 
$\frac{\Delta}{J}$ insures that $B_1\ll A_2$ and $A_1\ll B_2$. This, in turn, suggests that anomalously high survival
probability is promoted by antisymmetric combination, $a_1-a_2$ (see Eq. \ref{new1}). Symmetric combination, on the other hand, yields a conventional survival probability with the exponent $\sim \frac{J^2}{v}$.
As a final remark, our treatment of the $2\times 2$ model was semiclassical. This approach is justified when the velocity $v$ is small enough, namely, the probability
Eq. \ref{main} should be much smaller than one. More rigorous derivation of Eq. \ref{main} based on the 
Schr{\"o}dinger equation in the absence of decay
can be found in  Ref. \onlinecite{Rajesh}.

\section*{Acknowledgements}
We would like to thank Nikolai Sinitsyn for useful discussions. R.K.M. would like to thank  the U.S. Department of Energy, Office of Science, Basic Energy
Sciences, Materials Sciences and Engineering Division, Condensed Matter Theory Program. R.K.M. was also supported by the Center for Nonlinear Studies.

\end{document}